\documentclass[letterpaper]{article}

\usepackage{fullpage,authblk}
\usepackage{amssymb,amsfonts,amsmath,enumerate}
\usepackage[bookmarks, colorlinks=true, plainpages = false, citecolor = green, urlcolor = blue, filecolor = blue]{hyperref}
\usepackage{graphicx,subfigure,bm,latexsym}
\usepackage{multirow}
\usepackage{appendix}
\usepackage{float}
\usepackage{multicol}

\DeclareMathOperator{\sign}{sign}

\begin{document}

\title{Informative Ranking of Stand Out Collections of Symptoms: \\A New Data-Driven Approach to\\ Identify the Strong Warning Signs of COVID 19}

\author[1,2]{Abd AlRahman AlMomani}
\author[1,2]{Erik Bollt}

\affil[1]{Department of Electrical and Computer Engineering, Clarkson University, Potsdam, NY 13699, USA}
\affil[2]{Clarkson Center for Complex Systems Science ($C^3S^2$), Potsdam, NY 13699, USA}
\date{}
\maketitle
\begin{abstract}
We develop here a data-driven approach for  disease recognition based on given symptoms, to be efficient tool for anomaly detection. In a clinical setting and when presented with a patient with a combination of traits, a doctor may wonder if a certain combination of symptoms may be especially predictive, such as the question, ``Are fevers more informative in women than men?" The answer to this question is, yes. We develop here a methodology to enumerate such questions, to learn what are the stronger warning signs when attempting to diagnose a disease, called Conditional Predictive Informativity, (CPI), whose ranking we call CPIR.  This simple to use process allows us to identify particularly informative combinations of symptoms and traits that may help medical field analysis in general, and possibly to become a new data-driven advised approach for individual medical diagnosis, as well as for broader public policy discussion.  In particular we have been motivated to develop this tool in the current environment of the pressing world  crisis due to the COVID 19 pandemic.  We apply the methods here to data collected from national, provincial, and municipal health reports, as  well as additional information from online, and then curated to an online publically available Github repository.

\end{abstract}

\section{Introduction}

As healthcare systems around the world copes with the COVID-19 crisis, concerns about the ongoing spread of the disease remain, in part because of an as yet uncertain size and spreading role concerning asymptomatic and low symptomatic population, and also the true false negative rate \cite{falsenegative} of available tests.  Such a clear threat to humanity demands our best informative  ability to recognize its traits across populations. Clinical diagnosis remains a major line of defense, especially while direct tests remain uncertain and not widely available.

\begin{figure}
    \centering
    \includegraphics[scale=0.4]{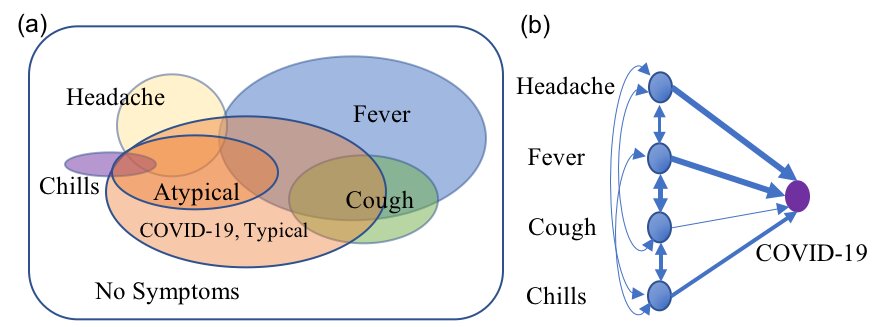}
    \caption{(Left) Venn diagram of symptoms, typical, and atypical sets of patients. This figure illustrates how some patients share different symptoms, which represent mutual information shared between different symptoms. (Right) Graph representation of symptoms that shows the shared information between symptoms and the disease, and between symptoms themselves. Our interest is to discover the \textit{direct} information between symptoms and the disease.}
    \label{fig:Venn}
\end{figure}

We develop in this work a Conditionally Predictively Informative Ranking (CPIR) method, that adopts data-driven principles of information theory, as an extension of the principle of causation entropy \cite{CSE}, to give a reliable ranking that reflects the direct informativity of each symptom, after considering the underlying relationships between the symptoms. See Fig.~\ref{fig:Venn}.  In brief, this approach contrasts directly to the commonly used concept of correlations \cite{kirkwood2010essential, bland2015introduction, james2013introduction, peck2015introduction}, that associates a score to describe how strongly measurements go together, whereas, CPIR associates a score to describe how strongly atypical is a new observation, given that a collection of symptoms has already been observed.  Often it is the presence of atypical observations that allow the clinician to decide how worrying should be considered the constellation of symptoms presenting in a given patient.  Likewise, understanding atypical associations can be especially informative when characterizing a disease across the population.

We define a Conditional Predictive Informativity (CPI) measure in terms of conditional mutual information  ($I$) to describe how much information is to be inferred if a given patient has certain symptoms, given already stated other symptoms are observed.  That is, in answering the question, ``Is a cough especially informative if a fever has already been observed in a patient," would be stated with the CPI as the following and relating to conditional mutual information,
\begin{center}
    $CPI$(cough associates to a patient if a fever is already observed) $=I$ (cough of a patient given fever),
\end{center}
referring to the available COVID-19 dataset, and as shown in Fig.~\ref{fig:Venn}, it turns out those with a fever is about 80\% of those already have a cough, so while correlation picks that association it is exactly not this association we wish to identify with CPI. 

Another example, that we may ask, ``Are chills especially informative if a headache has already been observed in a patient?'' For that question,  we write:
\begin{center}
    $CPI$(chills associates to a patient if a headache is already observed) $=I$ (chills of a patient given headache).
\end{center}
Referring to the available COVID-19 dataset, we find that less than 20\% of patients who have chills  also have a headache.  So the answer to the question is yes, observing chills is informative given a headache. Moreover, chills is crucially informative. As we show in Fig.~\ref{fig:Venn}, patients can be seen as typical and atypical patients. Both groups have tested positive for the disease, however, while the majority of the patients shares the same set of symptoms and reaction to the disease, a subgroup will have different reactions that may appear as severe pain and symptoms, and even death. In COVID-19 statistics, we see that more than 80\% of the patients have mild symptoms, or they may even be asymptomatic, while a small group have severe pain and risky symptoms, and the global mortality rate of COVID-19 is less than 7\% of confirmed cases. 

Even while the  number of patients who have chills is much lower than patients who have fever, see in Fig.~\ref{fig:Venn}, this symptom turns out to be more informative than fever, meaning that chills  specially appears in atypical patients, but not  in typical patients.  However, a fever appears in both groups, and furthermore, it is highly associated with other diseases.  This paper is devoted to developing and describing how to compute these kinds of classifications and conclusions.

\begin{figure}
    \centering
    \includegraphics[scale=0.4]{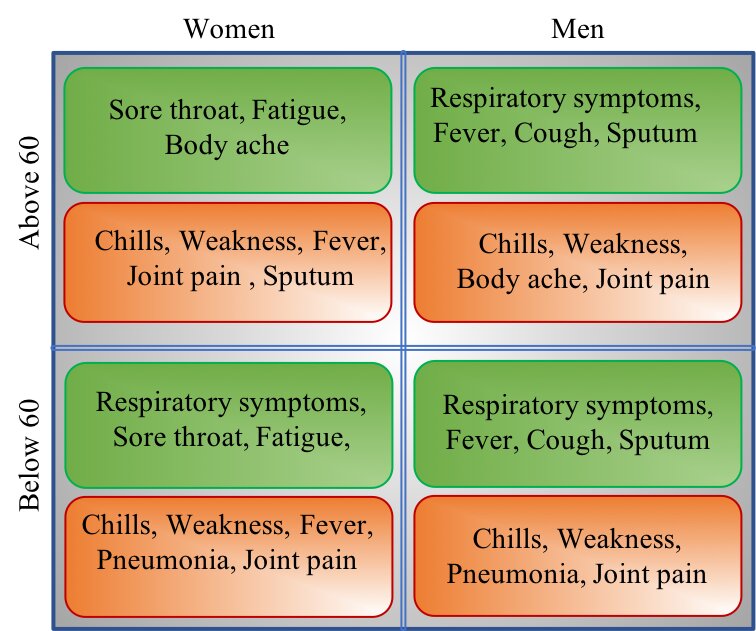}
    \caption{Based on the results in Fig~\ref{fig:CPIR}-Fig.~\ref{fig:age}, we summarize the outcome of exaustive 
     search of combinations of demographic information and symptoms that maximize the CPI, the most striking of which we highlight here. In green boxes, we show the symptoms associated with typical patients, and in red, we show the critical and risky symptoms that are associated with the atypical set of patients.}
    \label{fig:summary}
\end{figure}

Ranking on this conditional measure allows us to learn the markers of the disease. To state one such example outcome, our analysis shows that chills combined with pneumonia is especially informative when it occurs in young women, more so than in other patients.  It is a standout collection of symptoms.  Another example is that headaches and body aches are highly informative of atypical patients when they appear in men. More results of this type are shown in Fig.~\ref{fig:summary}.  With such examples in mind, our methods here will allow us a data-driven method to  define an informativity ranking of symptoms toward a reliable prediction of disease presence based on the specific combination of informative symptoms.  The knowledge gained when observing a specific combination of symptoms relative to the further predictability of the status of other symptoms founds the essential ability to understand how to understand observations in the clinical setting.

With the idea of informative collections of symptoms, then we offer the CPIR, as a ranking of these informative symptoms, sorted so as to describe those symptoms that when occurring together, make for unmistakable signs of the disease. Stated another way in terms of an example, a patient who is observed to have a dry cough and a fever is not so indicative of the disease since so often the fever and a dry cough go together. Just as even more so in the extreme, observing a dry cough and then the presence of a left foot, then furthermore observing the presence of a right foot is not so informative as a further observation.  However, as it turns out we find, a cough may be indicative, but then a fever may be more informative if furthermore the patient is female. 

Individual-level epidemiological data from the COVID-19 outbreak, are publically available from the international resource \cite{data}.  These are collected from national, provincial, and municipal health reports, as well as additional information from online, and these data are collectively curated for general use. This data is continually updated, as described in the paper associated with the Github repository. As of April 16, the dataset has more than 267,000 entries for individual-level data.  Fig.~\ref{fig:summary} is a table of the more striking combinations of informative symptoms and traits, that we have found in this study, the methods of which to rank the CPI are developed below.

\section{Methods}
In our previous work, \cite{Almomani2020}, we introduced the method of Entropic Regression, which adopts the principle of causation entropy, \cite{CSE}, for the discovery of underlying dynamics based on the influence of a set of candidate functions to the outcome dynamics. However, in the case of Boolean outcomes, and mixed datasets, where the outcome depends on a combination of variables that could be real numbers, Boolean variables, or descriptive data, we must extend principles of causality inference and minimal description.

\subsection{\textit{``Booleanization''} of Non-Boolean Variables}

\begin{figure}[h!]
    \centering
    \includegraphics[scale=0.4]{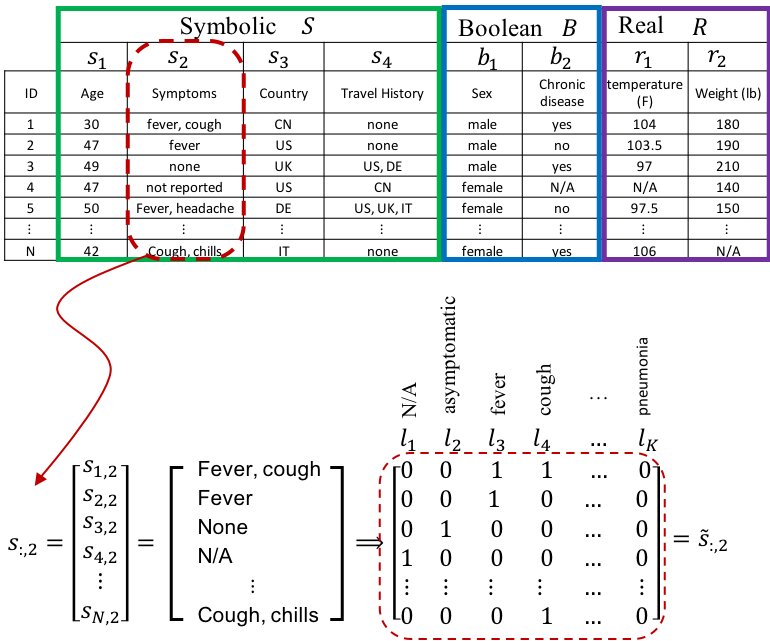}
    \caption{Booleanization of non-Boolean variables illustrative example. By this example, we illustrate how to convert standard descriptive variables, such as the commonly named symptoms, into a Boolean matrix as this is a required initial stage for our analysis.   See Eqs.~(\ref{step1})-(\ref{step3}).}
    \label{fig:excel}
\end{figure}

Let $Y \in \{0,1\}$ be the outcome that a patient has a disease, or not, such that:
\begin{equation}\label{step1}
    Y = \begin{cases} 0, \text{ if the disease test is negative,} \\ 1, \text{ if the disease test is positive.} \\  \end{cases}
\end{equation}
The patient data, or the variables that influence the outcome, are mostly mixed variables with different data types. See Fig.~\ref{fig:excel}. Further, define the $i^{th}$ patient variables
to be, \begin{equation}\label{step2}
    \mathcal{V}_i = [s_{i,1},s_{i,2},\dots,s_{i,n_s}, b_{i,1},b_{i,2},\dots,b_{n_b}, r_{i,1}, r_{i,2}, \dots, r_{n_r}],
\end{equation}
(i.e. a row in the table shown in Fig.~\ref{fig:excel}).  Here, $\{s_{i,j}\}_{j=1}^{n_s}$ are symbolic variables that each  has independent sets of labels (i.e. symbols, category,...,etc.) $\{l_1, l_2, \dots, l_{K_j}\}$, and $n_s$ is the number of symbolic variables. Similarly, $\{b_{i,j}\}_{j=1}^{n_b} \in \{0,1\}$ are Boolean variables, and  $n_b$ enumerates the Boolean variables.  $\{r_{i,j}\}_{j=1}^{n_r}$ are real valued variables, and $n_r$ is number of real valued variables.\\

For example, and as preliminary introduction to our approach, assume that the variable $s_{:,2}$ is an $N$-dimensional vector, representing the number of observations or diagnosis states of symptoms of a given patient.  Symptoms are labeled;  for example  $l_1$ represents  presence of a fever, $l_2$ represents  presence of a cough, and $l_3$ represents a headache, or possibly even the degree of these. Here we illustrate with example that $K_2=3$ is the number of labeled descriptions, or the number of symptoms that will be observed in patients. For sake of clarity we will drop the subscript and we write the vector of symptoms for all patients as $s$. For a given patient $i$, who for example has a fever and a headache, we write the entry $s_{i} = \{l_1,l_3\}$, representing occurrence of a fever and a headache. 
Then, we can translate the vector descriptive variable $s$, to multi-vector (matrix, see Fig.~\ref{fig:excel}) Boolean variables $\Tilde{s} = \mathcal{B}(s)$, such that:
\begin{equation}\label{step3}
    \Tilde{s}_{i,k} = \mathcal{B}(s_i) = \begin{cases} 1,  \text{ if } l_k \in s_i \\ 0, \text{ otherwise}. \end{cases}
\end{equation}
So, the function $\mathcal{B}(s)$ converts the column vector $s$ of $N$ descriptive entries, to an $N\times K$ Boolean matrix $S$, where the $k^{th}$ column, $k=1,...,K$, of the Boolean matrix describes the occurrence of the labeled description $l_k$ on $s$. Fig.~\ref{fig:excel} shows a schematic illustration for this local ``Booleanization'' process.  With variables Booleanized, \textcolor{black}{we now describe in the next section how to construct a conditionally predictively informative ranking (CPIR) for the symptoms.}

It is clear that real valued variables can be converted to a set of Boolean variables by thresholding and then these translated to labels and then to a Boolean matrix as discussed above. 

\subsection{Conditionally Predictive Information} 

Now we describe how certain Boolean factors may be predictive of other factors, given the status of yet other factors. 
This is a variation on the theme of our previous work in causation entropy \cite{cafaro2015causation, CSE, sun2014identifying, sun2015causal}, but here given the nature of the expected data and the corresponding difference of the underlying questions, the role of time is less relevant than the notion of indicative and predictive.  Our goal here is to define an informativity ranking of symptoms toward reliable prediction of disease presence based on the combination of informative symptoms. 

The Conditionally Predictive Information (CPI) described here as the conditional mutual information between one Boolean variable $S_k$ and the outcome $Y$ is given by: 
\begin{eqnarray}\label{eq:cse}
    CPI_{S_k|S_{k^*} \rightarrow Y} & = & I(S_k ; Y | S_{k^*})
\end{eqnarray}
where $k^*$ is the set of Boolean variables indices $\{1,...,K\} - \{k\}$. That is the mutual information between the $k^{th}$ Boolean variable and the outcome $Y$, conditioned on all other Boolean variables.  While this is therefore just a conditional mutual information \cite{cover}, it is the way the conditioning set is designed, and how these are ranked that make this approach especially relevant to our needs here. Thus, it is a measure of knowledge one gains in observing some of the variables to  further the possibility to predict the status of other variables, which is a crucial when the goal is to uncover a minimal and most informative set of variables.

\begin{figure}
    \centering
    \includegraphics[scale=0.4]{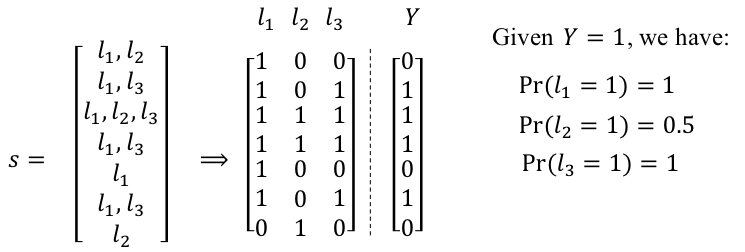}
    \caption{In a statistical sense, we see that $l_1$ and $l_3$ have the same probability of appearing when the outcome $Y=1$. However, we see that the variable $l_3$ has the exact state as $Y$. Meaning that if we consider $Y=l_3$ as our prediction map, we will have perfect accuracy, regardless of the values of $l_1$ and $l_2$. Then, we see that although $l_1$ and $l_3$ have the same probability, $l_3$ is more informative variable, and in the sense of conditional mutual information, we will see that $I(Y;l_k | l_3) = 0$, for all $k\neq 3$. That is, no variable has additional information other than the information provided by $l_3$.}
    \label{fig:CorrVsCaus}
\end{figure}

Eq.~\ref{eq:cse} quantifies the information added by observation of the Boolean variable $S_{k}$, given the state of all other symptoms.  However, this does not inform the directionionality of influence between $S_{k}$ and the outcome $Y$. For example, the Boolean variables $a = [1 1 1 0 0 0]$ and $b=[0 0 0 1 1 1]$ have high mutual information with the outcome $y =[1 1 1 0 0 0]$. However, while the event occurrence in $y$ is associated  with the event occurrence in $a$.  We see that the event occurrence in $y$ is associated with non-occurrence of the event in $b$. Then, the mutual information $I(a;y) = I(b;y)$ may result, but this reveals no information about the directionality of the relationship. To address this issue, we re-write Eq.~\ref{eq:cse} as:
\begin{eqnarray}\label{eq:cse2}
    CPI_{S_k|S_{k^*} \rightarrow Y} & = & \Gamma(S_k,Y) I(S_k ; Y | S_{k^*})
\end{eqnarray}
where $\Gamma$ is a sign function that is given by:
\begin{eqnarray}
    \Gamma(S_k,Y) & = & \sign\left( Pr(Y | S_k) - Pr(Y | ^\lnot S_k) \right) \notag\\
                  & = & \begin{cases} 1, Pr(Y | S_k) > Pr(Y | ^\lnot S_k) \\ 0, Pr(Y | S_k) = Pr(Y | ^\lnot S_k) \\ -1, Pr(Y | S_k) < Pr(Y | ^\lnot S_k) \end{cases} 
\end{eqnarray}
where $Pr(Y | S_k)$ is the conditional probability of $Y$ given $S_k$, and ($^\lnot$) is the logical NOT operator of a Boolean variable. That is, if the occurrence of ($Y=1$) associates with occurrence of ($S_k=1$) more than it is associates with $S_k = 0$, then this describes a positive relationship. Otherwise it is negative relationship, that occurrence of $Y$ is associated with non-occurrence of $S_k$.

Here, we want to emphasize a major difference between our approach and the commonly used methods from statistical analysis by correlations from data. Statistical analysis based on correlations \cite{kirkwood2010essential, bland2015introduction, james2013introduction, peck2015introduction} has a different goal than we have here. Correlations describe variables and outcomes in terms of probability of occurrence, but it is well known that correlation does not imply causation \cite{corr}. For example, suppose that 80\% of the patients ($Y=1$) have a fever, and 50\% of them have a cough. Then, fever and cough are the dominant symptoms. However, this analysis neglects the correlation among the symptoms themselves,  but most important to our interests here, it does not reveal if the symptoms are informative or not. See Fig.~\ref{fig:CorrVsCaus} for a detailed example. We are less interested in the cause in this work, and instead, we are interested in developing an Informativity Rank (IR) of the symptoms. That is, for a set of symptoms, what are the symptoms that are Conditionally Predictively Informative (CPI) of an underlying condition such as the presence of a disease.

Assume an example in the extreme for discussion, that 99\% of the patients who have a cough also have the fever. Then clearly there is (almost) no further useful information provided by observing the cough that was not already provided by observing the fever. Thus, the cough on its own can not stand as an informative symptom if not combined with the fever. Seeking informative variables is our main objective, and the conditional mutual information, carefully conditioned as stated and with the optimized conditioning set, can reveal this informative set by discovering the set of direct influences between the variables and the outcomes. Finding the conditional mutual information in Eq.~\ref{eq:cse}.  In other words, this translates to asking, ``What is the \textit{direct} influence (information) provided by the symptom $S_k$, given the information from all other symptoms, $S_{k^*}$?'' 
By this local analysis, we will infer the dominant variables are most informative. Furthermore, we will show this analysis  in terms of the COVID-19 dataset discussed above. First, however we must explain an important issue associated with uni-state outcomes.

\subsection{Uni-State Outcome Variable} 
In the available COVID-19  data, we must cope with a common and well  known problem called selection bias, \cite{selec}; this problem realizes itself here in that we generally only have the data acquired from actual diseased patients, meaning that we only have the data for the people who have already tested positive.  We call this the problem of a uni-state outcome variable, $Y = 1$, for all the patients.

Considering the COVID-19 dataset, we found about 26 commonly reported symptoms (fever, cough, ..., etc.), including the asymptomatic cases to be within the possible combinations of symptoms. This means that for each patient $i$, we have a binary string $S_i$ of 26-bits that represents the symptoms that patients have.
For simplicity of discussion, we will consider an example of 4 symptoms ordered as follows: \{Asymptomatic, Fever, Cough, Abdominal pain\}, but of course we use all the available data in our analysis. Further, if certain of the symptoms for the patient are not reported, then we have $S_i = [0,0,0,0]$.  If it is reported that the patient has no symptoms, we have $S_i = [1,0,0,0]$. If the patient has a fever and a cough, we have $S_i = [0,1,1,0]$, and so on.

\begin{figure}
    \centering
    \includegraphics[scale=0.5]{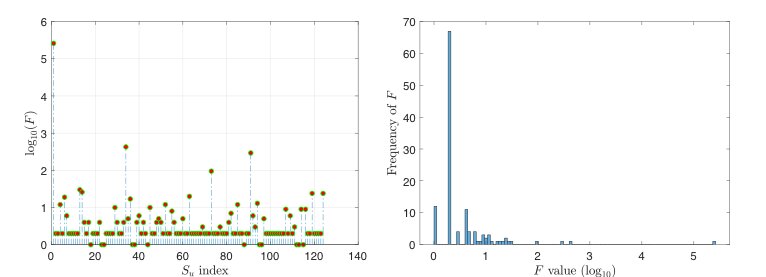}
    \caption{(Left) Frequency (Histogram) of the unique symptoms combination $S_u$. (Right) Frequency of the Frequency (Hist-Histogram). We see to the left that most of the unique symptoms combinations have very low occurrence in the dataset, and to clarify that, we consider the histogram of the frequencies to the right, which shows that most of the frequencies have low values, which means that the majority of symptoms combination $S_u$, have only one or two occurrence in the 260000 entries of the dataset.}
    \label{fig:hist}
\end{figure}

For a 26-bit binary string, there exists the possibility of up to $2^{26}$ or more than 67 million possible combinations of symptoms. However, we found that in the actual data set, there were expressed in fact only $n_u = 124$ unique binary strings $S_u$ for the actual symptoms observed. In Fig.~\ref{fig:hist}, we show the histogram of $S_u$ derived from the data. Observe that not only do we have a low count of unique states, moreover, most of the unique states have a very low frequency of occurrence $F = 1, F = 2$. Low frequency events, associated with a low probability or otherwise rare events, may be especially interesting to be studied individually as these outliers may in fact be specifically informative. However, our intention here is to rank the symptoms based on their informativity of the disease on the majority of patients.

\begin{figure}
    \centering
    \includegraphics[scale=0.4]{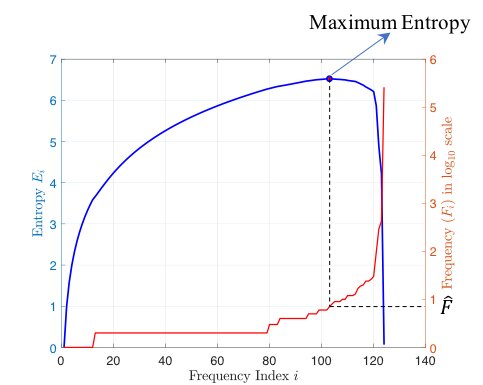}
    \caption{Entropy of the unique states. We see that as we adding states with low frequency, the entropy increase. However, at some point where the frequency of the state is significantly larger than previous frequency, the entropy decrease. The point of maximum entropy can be seen as a critical point, and the states before and after this point have different information and belong to different categories.}
    \label{fig:maxEntropy}
\end{figure}

Note that we seeks to infer the CPI that describes the outcome of most patients, and hence, we assume that the outcome $Y$, is equal to 0, if the patient has a binary string $S_i$ that has low probability. Mathematically we write:
\begin{equation} \label{eq:Ygenerating}
    Y_i = \begin{cases} 1, Pr(S_i) > \delta \\ 0, Pr(S_i) \leq \delta \end{cases}
\end{equation}
where $Pr(S_i)$, is the probability of occurrence of the binary string $S_i$, and $\delta$ is a probability tolerance. In order to choose $\delta$, assume that $F$ is the ordered frequency (histogram) of the unique set of symptoms $S_u$, such that $F_1 \leq F_2 \leq \dots F_{n_u}$. To investigate how each state in $S_u$ provides additional information, we consider to analyze the entropy of the histogram sequentially. Let the entropy $E_i$ be the entropy of the probability distribution of the frequency entries, given by:
\begin{equation}\label{eq:maxEntropy}
    E_i = \sum_{j=1}^{i} \frac{F_j}{\sum_{k=1}^{i} F_k}\log_{2}(\frac{F_j}{\sum_{k=1}^{i} F_k}) 
\end{equation}
where $\frac{F_j}{\sum_{k=1}^{i} F_k}$ represents the probability of the $j^{th}$ entry in $F$ with respect to  the assumption that only the first $i$ states are available. In another words, at each $i$ we are asking what is the entropy of $S_u$ if we were to assume that we have  only have the first $i$ states?. This allow us to track, starting from the low frequency states, how the large frequency states affects the information (entropy) of $S_u$. Fig.~\ref{fig:maxEntropy} shows the entropy curve from Eq.~\ref{eq:maxEntropy}.

It will be a subject for our future work to connect this approach to the theory stemming from the asymptotic equipartition property (AEP) \cite{cover}, and in so doing, to discuss the optimality of the value of $\delta$, whereby in analogy to AEP, we are classifying the patients as associated with typical and atypical sets.  Thus, most of the information is interpreted to be  associated with the typical set of patients. For our current analysis, we consider $\delta = \hat{F}/N$, where $\hat{F}$ is the frequency at the maximum entropy, and $N$ is the sample size. For our dataset, we found $\delta = 2.69 \times 10^{-5}$.

Now, given $\delta$, we have the outcome $Y$ from Eq.~\ref{eq:Ygenerating}, and the set of Boolean variables $S$ that describes the symptoms. We apply Eq.~\ref{eq:cse2} to find the conditional mutual information (CMI) of each symptom. The CPIR is the scaled CMI, $\sim [0,1]$, to indicate the rank of each symptom and it is shown in Fig.~\ref{fig:CPIR}. In Fig.~\ref{fig:symptomsNetwork}, we show the CMI between the symptoms themselves, which are the CMI between each pair of symptoms, given the information from all others. This mutual information in Fig.~\ref{fig:symptomsNetwork} between the symptoms, can practically lead to a fuzzy classification of the critical symptoms since it is hidden for the commonly used statistical techniques. 

\subsection{\textcolor{black}{Computational Approach}}
The conditional mutual information associated with the CPI, Eq.~(\ref{eq:cse}), requires that we review the conditional mutual information \cite{cover}, and remark how the associated probabilities may be estimated.
Let $X, Y$, and $Z$ be jointly distributed random variables with associated probability density function $p(x,y,z)$. The conditional mutual information between $X$,$Y$ given $Z$ according to Eq.~\ref{eq:cse}, is given by:
\begin{eqnarray}\label{eq:cmi}
    I(X;Y | Z) & = & -\sum_{x,y,z} p(x,y,z)\log\left( \frac{p(x,y|z)}{p(x|z)p(y|z)} \right) \notag \\
               & = & H(X|Z) - H(X|YZ) \notag \\
               & = & H(XZ) + H(YZ) - H(XYZ) - H(Z),
\end{eqnarray}
where $H$ is the entropy function, and $H(\cdot\cdot)$ is the joint entropy of two variables. The entropy of a discrete random variable $X$ with pmf $p_X(x)$, and $n$ possible states is given by
\begin{equation}\label{eq:entro}
    H(X) = -\sum_{i=1}^{n} p(x_i)\log(p(x_i)).
\end{equation}
For a Boolean random variable $X$, let $\hat{p}$ be the probability that $X=1$.  We do not emphasize here how to efficiently estimate $\hat{p}$ other than to note that typically this may be done by counts of relative occurence. Then, the probability that $X=0$ is $1-\hat{p}$, and the entropy of $X$ is then given by:
\begin{equation}
    H(X) =  -\hat{p}\log(\hat{p}) - (1-\hat{p})\log(1-\hat{p}).
\end{equation}

For the joint entropy of discrete variables in Eq.~\ref{eq:cmi}, we can think of the joint variable $XZ$, for example, as a new variable with a two dimensional outcome state (concatenation of the two variables as two columns), for which we call the joint space of $X$ and $Z$. Then, the entropy can be found by the probability distribution of the unique states in $XZ$ (unique rows), using Eq.~\ref{eq:entro}.

Now, we can algorithmically obtain our conditionally predictively informative ranking (CPIR) by the following loop through all symptoms:
\begin{equation}
    \begin{cases} \Gamma_i = \Gamma(S_i,Y) \\ CPI_i = \Gamma_i I(S_i;Y|S_R) \end{cases},
\end{equation}
and the ranking CPIR is then given by $CPIR_i = CIP_i / \max(\vert CPI \vert)$. In Fig.~\ref{fig:CPIR}, we show the results with plotting the absolute value of CPIR after descending sorting, which gives more clear view and readability for the figure.

\section{Results, Symptoms  and Informativity Questions}

Now we are in a position to answer very simple but important questions, such as, ``Are fevers more informative in women than men?"  Clearly there are many comparably, important, and simple to state questions that become clinically relevant when a doctor may be presented with a specific patient presenting specific symptoms. One interesting question to ask is that if the symptoms have different CPIR depending of specific demographic variable such as sex or age. 
There are different ways to answer this question with the conditional mutual information. We may for example,  add the Boolean vector of gender to the conditioning set, and track the reduction of CPIR for each symptom. However, another approach that overcomes the need to increase the size of the conditioning set is to replace the outcome set with the gender vector.

\begin{figure}[h!]
    \centering
    \includegraphics[scale=0.4]{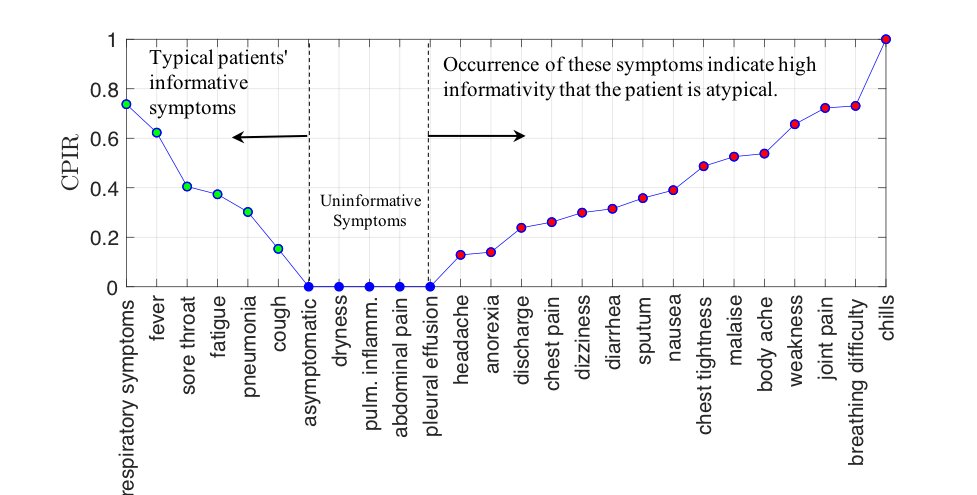}
    \caption{Conditionally Predictively Informative Ranking (CPIR). Although statistically, fever is the highest occurrence symptom, several other symptoms was more informative than fever.  The most reliable five informative symptoms are: chills, breathing difficulty, respiratory symptoms, joint pain, and fever. To understand how to read this curve,  consider the cough, which is statistically known to be from the main symptoms of COVID-19. However, we see here that the cough is a weak informative symptom, meaning that the cough by itself is not a reliable indication of COVID-19 unless it jointed with a stronger symptom such as fever. See Fig.~\ref{fig:Venn} for more details. On the other hand, if a patient has only chills, that is a more reliable indication than cough or fever alone. We can conclude then that the high informative symptoms jointly, can be a solid indication that the patient has COVID-19, such as having headache and sputum together, or headache and fatigue together, and so on. Moreover, symptoms appear to the right (red dots), are more critical, since they indicate informative symptoms of atypical set of patients.
    }
    \label{fig:CPIR}
\end{figure}

\begin{figure}[h!]
    \centering
    \includegraphics[scale=0.4]{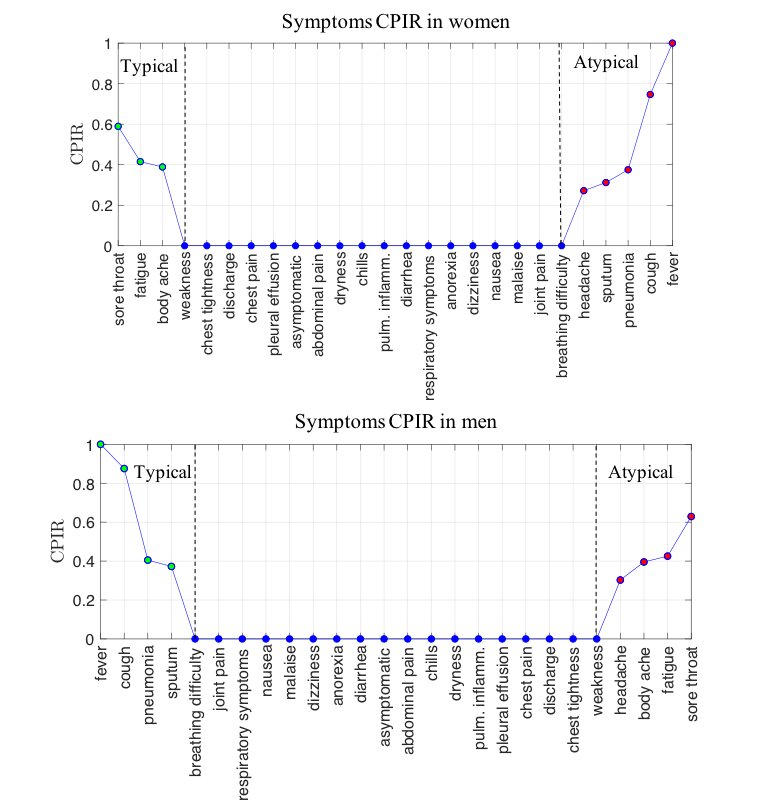}
    \caption{Symptoms Demographic Informativity (Gender). It is important to note that this figure is independent of the order and analysis that are shown in Fig.~\ref{fig:CPIR}, since here we have a different outcome, and we are investigating the mutual information between symptoms and demographic variables. Low (or zero) CPIR in this figure does not mean that the symptom is not informative, but it means that the symptom has no different (bias) informativity between women and men. (Top) We see the results for symptoms informativity bias in women, and we see the dominant informativity in this figure is that fever and cough, are informative for the atypical set of patients in women, while they are associated with the typical set in Fig.~\ref{fig:CPIR}. This indicates that fever and cough are more surprising to appear in women. Then, they are more critical in women, where their presence in women increases the probability that the patient is from the atypical set. (Bottom) Similarly, we see the results for symptoms informativity bias in men, and we see the dominant informativity in this figure is that fever and cough are associated with the typical set of patients. Although that reduces the risk in men who have fever and cough, however, since 75\% of patients have a fever, then that indicates that men have a higher probability of infection with COVID-19. We note that our dataset has some missing or unreported information about gender, and that is why we see the difference between the graphs. If we have complete data about the gender, with only male and female entries, then we will see the CPIR for men as rotation (left-right) of the women graph, and that can be seen clearly in Fig.~\ref{fig:age}.}
    \label{fig:sexage}
\end{figure}

Let, $Y_{i,1} = 1$, if the patient $i$ is female, and $Y_{i,1} = 0$ otherwise. Similarly, let $Y_{i,2} = 1$, if the patient $i$ is male, and $Y_{i,2} = 0$ otherwise. Note that due to missing data and other factors, $Y_{:,1} \neq ^\lnot Y_{:,2}$, where ($\lnot$) is the logical operator (NOT). Then, we repeat the process for $Y_{:,1}, Y_{:,2}$, by using the symptoms matrix $S$, and Eq.~\ref{eq:cse2}. Fig.~\ref{fig:sexage} shows the results of symptoms demographic informativity.
Given $\delta$, we have the outcome $Y$ from Eq.~\ref{eq:Ygenerating}, and the set of Boolean variables $S$ that describes the symptoms. We apply Eq.~\ref{eq:cse2} to find the CPI of each symptom. The CPIR is the scaled CPI, as discussed in the computations section, to indicate the rank of each symptom, and it is shown in Fig.~\ref{fig:CPIR}. 

The fever and cough are widely known as the main symptoms of COVID-19, and in our dataset, we found that 75\% of patients have a fever, and 45\% of them have a cough. However, Fig.\ref{fig:CPIR} shows that the most informative symptom is chills, and it is specifically informative that the patient is from the atypical set of patients. The fever, placed the second informative symptom of the typical patients, while the most informative symptom of the typical patients was the respiratory symptoms, which include respiratory infection, acute respiratory viral infection (ARVI), and acute respiratory distress syndrome (ARDS). Clinically, breathing difficulty can be listed under respiratory symptoms. Analyzing breathing difficulty individually,  we found it to be the second informative symptom of atypical patients.

Clinically, fatigue and weakness are two different symptoms, where the weakness is defined as a failure to generate the
required or expected force on first testing or attempted performance, and fatigue is defined as a failure to generate the
required or expected force during sustained or repeated contraction \cite{edwards1978physiological}. Both symptoms are informative symptoms of COVID-19, however, weakness is more critical, since it is informative of atypical patients. Interestingly, since they are often mistakenly used alternatively, especially by the patients when they describe them, if we consider them to be equivalent, say both of them are fatigue, then it will be the most informative symptom among all of the other symptoms of COVID-19. 

\begin{figure}[h!]
    \centering
    \includegraphics[scale=0.4]{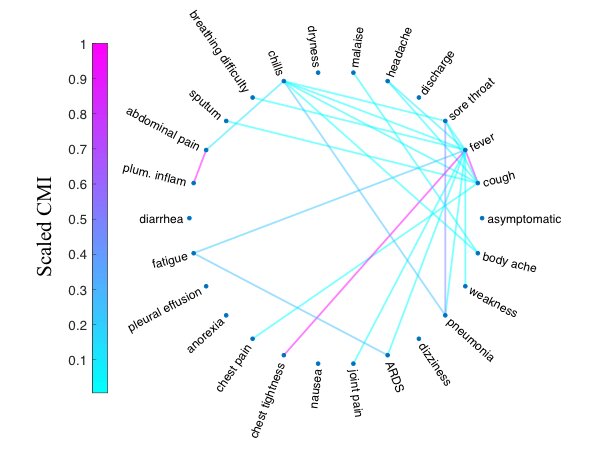}
    \caption{Symptoms influence network. In this figure, we show the mutual information directly shared between symptoms. We see that some symptoms have high mutual information between them, which makes the additional information provided by one of them, given the other, is low.}
    \label{fig:symptomsNetwork}
\end{figure}

In Fig.~\ref{fig:symptomsNetwork}, we show the CMI computed between pairs of symptoms, given the information from all others. This mutual information in Fig.~\ref{fig:symptomsNetwork} between the symptoms, can practically lead to a fuzzy classification of the critical symptoms since it is hidden for the commonly used statistical techniques. For example, in a statistical sense, we say that fever and cough are the most symptoms to appear in COVID-19 patients, and this analysis lack any consideration of the dependency between cough and fever, and that most of the patients who have cough are already have a fever. Fig.~\ref{fig:symptomsNetwork} shows this dependency as high direct information shared between cough and fever, and in our CPI approach, addressing these interactions between symptoms is embedded in Eq.~\ref{eq:cse2}, and we obtain the direct informativity between each individual symptom and the outcome.

To the question, are fevers more informative in women than men? The answer, according to the CPIR, is yes. 
In Fig.~\ref{fig:sexage}, we see that while fever is informative for the typical male patients, it is associated with the atypical female patients. In order to read Fig.~\ref{fig:sexage} fully and correctly, we say that fever is an informative symptom of atypical patients of women, which means that it is more surprising to see women who have a fever, and hence, it can be a more critical or risky symptom to be taken especially seriously when it appears in women. 

We must interpret that Fig.~\ref{fig:sexage}  helps  in recognizing differences in symptoms informativity between men and women, but it does not replace the general audience discussion of Fig.~\ref{fig:CPIR}, but rather it compiments it. For example we see that chills, which is the most informative symptom in Fig.~\ref{fig:CPIR}, has zero CPRI in men and women but that does not mean it is not informative.  Instead it means that chills have no difference in informativity between men and women. 
Finally we show in analogy to Fig.~\ref{fig:sexage}, a comparable assay of results for symptoms informativity in younger and older patients. 

As a summary of our results querying for the most striking combination of symptoms and traits, see a tabular summary Fig.~\ref{fig:summary}.

\begin{figure}
    \centering
    \includegraphics[scale=0.4]{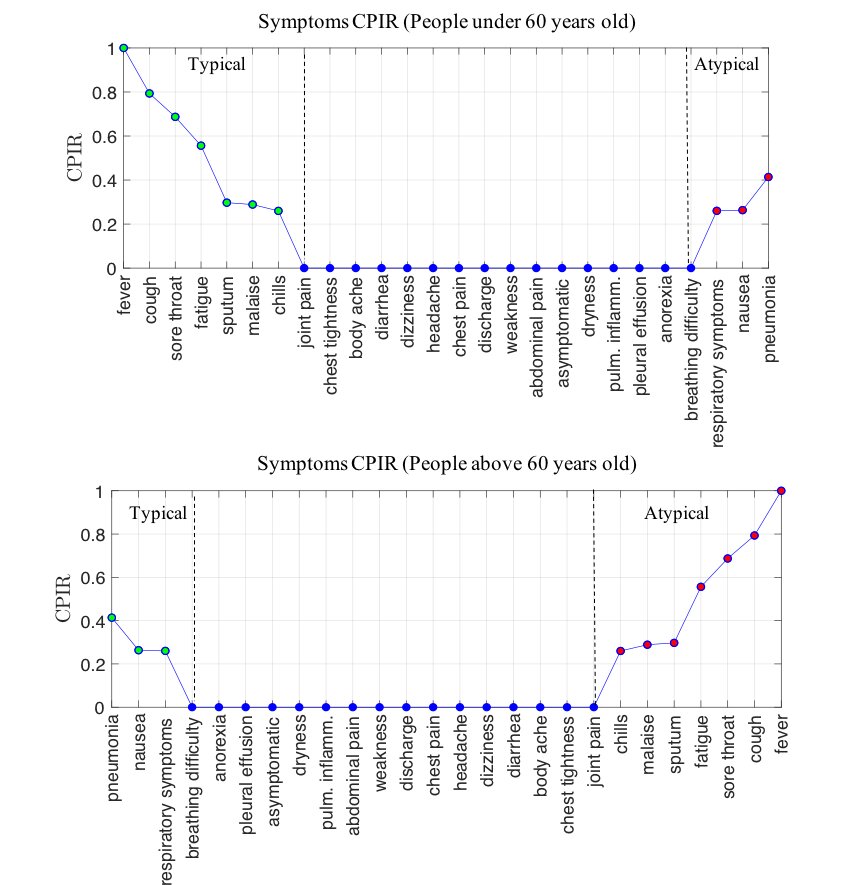}
    \caption{Symptoms Demographic Informativity (Age). We see the results for symptoms informativity bias based on the age of the patients. We see that fever, cough, sore throat, and fatigue are more informative in younger patients ($<60$), while pneumonia is more informative in older people.}
    \label{fig:age}
\end{figure}

\section{Discussion}

In this paper we introduced a Conditionally Predictively Informative Ranking (CPIR) approach. In partciular we used this method to analyze COVID-19 symptoms, and to give a  ranking of informative symptoms. In analogy to the symptoms example, other descriptive (labeled) variables can be analyzed to extract  the informative descriptions in each variable that contains labeled descriptions. In our future work, we will extend the idea to consider  symptoms-disease causality-driven networks, to construct informative networks that can give a signature of symptoms combinations.  We hope this can be helpful as a data-driven approach for  disease recognition based on given symptoms, and it can be efficient tool for anomaly detection.

From the presented mathematical methods applied to the data, the results are as shown in Fig.~\ref{fig:CPIR}, Fig.~\ref{fig:sexage} and Fig.~\ref{fig:age}.  Summarizing from these figures, we highlight in Fig.~\ref{fig:summary} our main perhaps most striking combination of symptoms and traits as our results.   While we have highlighted the COVID 19 in this discussion, for the obvious critical nature of this crisis, it is our hope that this tool, in particular the CPIR, may find utility for other disease analysis, and indeed for other questions of medical, social, and scientific importance.

\section{Code and Data Availability}

Our results are based on the dataset \cite{data}, which is continually updated online, and we will update our results based on the new data available, and the results will be updated continually in our online  \href{https://github.com/almomaa/COVID19}{(COVID-19 repository)}, together with the Matlab code to process the data and perform CPIR analysis.

\section{Acknowledgments}
This work was funded in part by the Army Research Office, and also DARPA.

\section{Conflict of Interest}
The authors declare no conflict of interest.

\bibliographystyle{plain}
\bibliography{covid19References}

\begin{thebibliography}{10}

\bibitem{Almomani2020}
Abd AlRahman~R AlMomani, Jie Sun, and Erik Bollt.
\newblock How entropic regression beats the outliers problem in nonlinear
  system identification.
\newblock {\em Chaos: An Interdisciplinary Journal of Nonlinear Science},
  30(1):013107, 2020.

\bibitem{bland2015introduction}
Martin Bland.
\newblock {\em An introduction to medical statistics}.
\newblock Oxford University Press (UK), 2015.

\bibitem{cafaro2015causation}
Carlo Cafaro, Warren~M Lord, Jie Sun, and Erik~M Bollt.
\newblock Causation entropy from symbolic representations of dynamical systems.
\newblock {\em Chaos: An interdisciplinary journal of nonlinear science},
  25(4):043106, 2015.

\bibitem{selec}
Corinna Cortes, Mehryar Mohri, Michael Riley, and Afshin Rostamizadeh.
\newblock Sample selection bias correction theory.
\newblock In {\em International conference on algorithmic learning theory},
  pages 38--53. Springer, 2008.

\bibitem{cover}
Thomas~M Cover and Joy~A Thomas.
\newblock {\em Elements of information theory}.
\newblock John Wiley \& Sons, 2012.

\bibitem{edwards1978physiological}
RHT Edwards.
\newblock Physiological analysis of skeletal muscle weakness and fatigue.
\newblock {\em Clinical science and molecular medicine}, 54(5):463--470, 1978.

\bibitem{james2013introduction}
Gareth James, Daniela Witten, Trevor Hastie, and Robert Tibshirani.
\newblock {\em An introduction to statistical learning}, volume 112.
\newblock Springer, 2013.

\bibitem{kirkwood2010essential}
Betty~R Kirkwood and Jonathan~AC Sterne.
\newblock {\em Essential medical statistics}.
\newblock John Wiley \& Sons, 2010.

\bibitem{corr}
Charles Ksir and Carl~L Hart.
\newblock Correlation still does not imply causation.
\newblock {\em The Lancet Psychiatry}, 3(5):401, 2016.

\bibitem{peck2015introduction}
Roxy Peck, Chris Olsen, and Jay~L Devore.
\newblock {\em Introduction to statistics and data analysis}.
\newblock Cengage Learning, 2015.

\bibitem{CSE}
Jie Sun and Erik~M Bollt.
\newblock Causation entropy identifies indirect influences, dominance of
  neighbors and anticipatory couplings.
\newblock {\em Physica D: Nonlinear Phenomena}, 267:49--57, 2014.

\bibitem{sun2014identifying}
Jie Sun, Carlo Cafaro, and Erik~M Bollt.
\newblock Identifying the coupling structure in complex systems through the
  optimal causation entropy principle.
\newblock {\em Entropy}, 16(6):3416--3433, 2014.

\bibitem{sun2015causal}
Jie Sun, Dane Taylor, and Erik~M Bollt.
\newblock Causal network inference by optimal causation entropy.
\newblock {\em SIAM Journal on Applied Dynamical Systems}, 14(1):73--106, 2015.

\bibitem{falsenegative}
Colin~P West.
\newblock Covid-19 testing: The threat of false-negative results.
\newblock In {\em Mayo Clinic Proceedings}, 2020.

\bibitem{data}
Bo~Xu, Bernardo Gutierrez, Sumiko Mekaru, Kara Sewalk, Lauren Goodwin, Alyssa
  Loskill, Emily~L Cohn, Yulin Hswen, Sarah~C Hill, Maria~M Cobo, et~al.
\newblock Epidemiological data from the covid-19 outbreak, real-time case
  information.
\newblock {\em Scientific Data}, 7(1):1--6, 2020.

\end{thebibliography}

\end{document}